\documentclass{article}
\usepackage{amsfonts}
\usepackage{amsmath}
\usepackage{enumerate}
\usepackage{hyperref}
\hypersetup{
    colorlinks=true,
    linkcolor=blue,
    filecolor=magenta,
    urlcolor=blue,
}

\usepackage{tikz}
\usepackage{tikz-3dplot}
\usetikzlibrary{matrix,arrows,positioning,decorations.markings,decorations.pathmorphing}

\newtheorem{theorem}{Theorem}

\newtheorem{proposition}[theorem]{Proposition}

\newcommand{\R}{\mathbb{R}}
\newcommand{\g}{\tilde{\mathfrak{g}}}
\newcommand{\sS}{\mathbb{S}}
\newcommand{\so}{\mathfrak{so}}

\begin{document}

\title{Rigid Body with Rotors and Reduction by Stages}
\author{Miguel \'{A}. Berbel \and Marco Castrill\'{o}n L\'{o}pez}
\date{}
\maketitle

\begin{abstract}
Rigid body with rotors is a widespread mechanical system modeled after the direct product $SO(3)\times \sS^1\times \sS^1 \times \sS^1$, which under mild assumptions is the symmetry group of the system. In this paper, the authors present and compare different Lagrangian reduction procedures: Euler-Poincar\'e reduction by the whole group and reduction by stages in different orders or using different connections. The exposition keeps track of the equivalence of equations as well as corresponding conservation laws.
\end{abstract}


\medskip


\section{Introduction}
The goal of this article is to give a complete and careful exposition of the Geometry and Mechanics of a rigid body with rotors when the groups of symmetries are brought on the scene. The symmetries of the problem are, as it is usually done in this context, employed to perform reduction of the phase space. In our case, we focus our attention to the Lagrangian formulation of the problem.

Rigid body with rotors is a recursive object in the literature and can be found in many references. We can refer to \cite{MScheurle} for a presentation in a pure theoretical point of view, but there are many others, a great quantity of them focused in applications (for example, some classical and recent publications are \cite{dissipation}, \cite{doroshin}, \cite{krisna}, \cite{hindi}, \cite{mircea} as well as the references therein). This research attention reveals the double importance of this model. On one hand, it serves as an excellent testing ground to implement the theoretical models developed in geometric reduction. On the other, the applications of the system itself are very valuable and can be found in many different situations, with particular emphasis in controllability.

Under mild assumptions, the group of symmetries of the rigid body with (three) rotors is the direct product $G=SO(3)\times \sS^1\times \sS^1 \times \sS^1$. The Lagrangian being invariant by the group $G$, the reduction procedure can be done directly with the full group (Euler-Poincar\'e reduction) to produce equivalent reduced equations. However, a reduction by the subgroup of the rigid body, that is, a reduction by the subgroup $SO(3)$ only, has deserved the main attention, separately by the reduction of the rest of terms $\sS^1\times \sS^1 \times \sS^1$. This splitting is consistent with the different nature of the symmetries, one dealing with the main body and the other with the rotors. This is a paradigmatic situation for which the theory of reduction by stages initiated in \cite{CMR} (see also \cite{LPcategory}) was built. But this chain of reduction can be done in different ways. Beginning with one subgroup or with the other. In addition, different possible choices of bundle connections appear along the way. This reduction process is scattered, abridged and incomplete in the literature, distributed in different papers. From this situation, a reference with all the details was in order. We provide in this work a unified, formal and comprehensive exposition for all the possible situations: we reduce by the full groups or by any subgroup, we keep track of the equivalence of the equations, and we analyze the corresponding conservations laws. A global presentation of these reductions fill a gap in the literature that will be of interest to future applications and models.

There are variants and different points of view of the problem of a rigid body with rotors. The Lagrangian for the rigid body subject to conservative forces is obtained subtracting a potential to the free rigid body Lagrangian. In what follows, only the free rigid body will be studied since that case is rich enough to explore the theory of reduction by stages and that perturbation is, conceptually, similar to the free case. Furthermore, the Hamiltonian side of the formulation (for example, the reader can have a look to \cite{moscow}) could be done easily. Finally, a non-holonomic variant of this problem is the spherical robot, where the rigid body considered is an homogeneous sphere rolling along a plane without slipping. ( for example, the control of this system is tackled in \cite{sphere}, \cite{kyushu}). The  symmetry group of this system is now the product $SO(2)\times \sS^1\times \sS^1 \times \sS^1$. Our presentation could be also applied to the spherical robot with the corresponding adaptation.

\section{Preliminaries}\label{preliminaries} \label{preliminaries}
Let $\bar{\rho}$ be a free and proper left action of a Lie Group $G$ on a manifold $Q$. The quotient $Q/G$ is a manifold and the projection $\pi_{Q/G,Q} :Q\rightarrow Q/G$ is a (left) principal $G$-bundle. We denote the infinitesimal generator of any $\xi\in\mathfrak{g}$, the Lie Algebra of $G$, at a point $q\in Q$ as
$$\xi ^{Q}_{q}=\left. \frac{d}{dt}\right\vert _{t=0}\exp (t\xi )\cdot q\in T_qQ.$$
A \textit{principal connection} $\mathcal{A}$ on $Q\rightarrow Q/G$ is a $\mathfrak{g}$-valued $1$-form on $Q$ such that
\begin{enumerate}[(i)]
\item $\mathcal{A}(\xi _{q}^{Q})=\xi$ for any $\xi \in\mathfrak{g}$ and $q\in Q$,
\item $\bar{\rho}_{g}^*\mathcal{A}=\mathrm{Ad}_{g}\circ \mathcal{A}$, for any $g\in G$,
\end{enumerate}
where $\bar{\rho}_{g}:Q\to Q$ denotes the action by $g\in G$ and $\mathrm{Ad}$ is the adjoint action of $G$ on $\mathfrak{g}$. At any $q\in Q$, a principal connection gives the splitting $T_{q}Q=H_{q}Q\oplus V_{q}Q$,
$$V_{q}Q=\ker T_q\pi_{Q/G,Q}=\{v\in T_{q}Q|T_q\pi_{Q/G,Q}(v)=0\},\qquad q\in Q,$$
$$H_{q}Q=\ker \mathcal{A}_{q}=\{v\in T_{q}Q|\mathcal{A}_{q}(v)=0\},\qquad q\in Q.$$
These are respectively called \textit{vertical} and \textit{horizontal} subspaces. The function $T_q\pi_{Q/G,Q}$ is an isomorphism between $H_{q}Q$ and $T_x(Q/G)$, where $x=\pi_{Q/G,Q}(q)$, the inverse of which $\mathrm{Hor}^{\mathcal{A}}_q$ is called \emph{horizontal lift}. In addition, the \textit{curvature} of a connection $\mathcal{A}$ is the $\mathfrak{g}$-valued $2$-form $$B(v,w)=d\mathcal{A}(\mathrm{Hor}(v),\mathrm{Hor}(w)),$$ where $v,w\in T_qQ$ and $\mathrm{Hor}$ denotes the projection from $ T_qQ$ to $H_{q}Q$.

The \textit{adjoint bundle} to $Q\to Q/G$ is the vector bundle $\tilde{\mathfrak{g}}=(Q\times \mathfrak{g})/G$ over $Q/G$ where the action of $G$ on $\mathfrak{g}$ is the adj int action. The elements of this bundle are denoted  by $[q,\xi]_{G}$, $q\in Q$, $\xi \in \mathfrak{g}$. There are three remarkable extra structures in $\tilde{\mathfrak{g}}\to Q/G$. Firstly, there is a fiberwise Lie bracket given by
$$\lbrack \lbrack q,\xi _{1}]_{G},[q,\xi _{2}]_{G}]=[q,[\xi _{1},\xi
_{2}]]_{G},\qquad \lbrack q,\xi _{1}]_{G},[q,\xi _{2}]_{G}\in
\tilde{\mathfrak{g}}_{x},\hspace{2mm}x=\pi_{Q/G,Q} (q).$$
Secondly, the principal connection $\mathcal{A}$ on $Q\to Q/G$ induces a linear connection, $\nabla^{\mathcal{A}}$, on the adjoint bundle given by the following covariant derivative along curves
$$\frac{D[q(t),\xi(t)]_G}{Dt}=\left[q(t),\dot{\xi}(t)-[\mathcal{A}(\dot{q}(t)),\xi(t)] \right]_G.$$
Lastly, there is a $\tilde{\mathfrak{g}}$-valued 2-form on $Q/G$ obtained from the curvature of $\mathcal{A}$ as
$$\tilde{B}(X,Y)=[q,B(\mathrm{Hor}^{\mathcal{A}}_qX,\mathrm{Hor}^{\mathcal{A}}_qY)]_G,$$
where $X,Y\in T_x(Q/G).$

Let $\rho:G\times V\to V$ be a free and proper action of a Lie Group $G$ on a vector bundle $V\to Q$ such that for every $g\in G$, $\rho_g:v\in V\to \rho(g,v)\in V$ is a vector bundle isomorphism. Then, $V/G$ and $Q/G$ are manifolds and there is a vector bundle structure on $V/G\to Q/G$ given by
$$[v_q]_G+[w_q]_G=[v_q+w_q]_G \text{ and } \lambda [v_q]_G=[\lambda v_q]_G,$$ where $[v_q]_G,[w_q]_G\in V/G$ are the equivalence classes of $v_q,w_q\in V_q$ and $\lambda \in \mathbb{R}$. A particular example of reduction of vector bundles is the tangent bundle of a principal bundle $Q$. This is utterly important in Mechanics when Lagrange--Poincar\'e reduction is performed on $G$-invariant Lagrangians defined on $TQ$.
In this case, the connection $\mathcal{A}$ induces a well-known vector bundle isomorphism:
\begin{eqnarray}
\alpha_{\mathcal{A}}:TQ/G &\longrightarrow  &T(Q/G)\oplus \g \label{identification}\\
\left[v_q \right]_G  & \mapsto & T_q\pi_{Q/G,Q}(v_q)\oplus [q,\mathcal{A}(v_q)]_G  \notag
\end{eqnarray}

Back to the case of an arbitrary vector bundle $V\to Q$, suppose that it has an affine connection $\nabla$ which is $G$-invariant, that is, $\nabla (\rho \circ s\circ \bar{\rho}^{-1})=\rho^{-1} \circ \nabla s\circ \bar{\rho}$, for any section $s\in \Gamma (V)$. We shall define a quotient connection on $V/G$. Let $X\in\Gamma (TQ/G)$, the identification \eqref{identification} gives a decomposition $X=Y\oplus \bar{\xi}\in \mathfrak{X}(Q/G)\oplus\Gamma(\g)\simeq \Gamma (TQ/G)$. Then, the unique $G$-invariant vector field $\bar{X}\in\Gamma^G(TQ)$ on $Q$ projecting to $X$ can be decomposed as $\bar{X}=Y^h\oplus W$ with $Y^h \in\mathfrak{X}(TQ)$ the horizontal lift of $Y$ and $W$ the unique vertical $G$-invariant vector field such that for all $x\in Q/G$, $\bar{\xi}(x)=[q,\mathcal{A}(W(q))]_G$ with $q\in \pi_{Q/G,Q}^{-1}(x)$. For $[v]_G\in \Gamma(Q/G,V/G)$ with $v\in\Gamma^G(Q,V)$ a $G$-invariant section, the \textit{quotient connection} is defined as
$$\left[ \nabla^{(\mathcal{A})}\right] _{G,Y\oplus \bar{\xi}}[v]_G=[\nabla_{\bar{X}}v]_G,$$
the \textit{horizontal quotient connection} is defined by
$$\left[ \nabla^{(\mathcal{A},H)}\right] _{G,Y\oplus \bar{\xi}}[v]_G=[\nabla_{Y^h}v]_G=[\nabla^{(\mathcal{A},H)}_{\bar{X}}v]_G$$
and the \textit{vertical quotient connection} is defined by
$$\left[ \nabla^{(\mathcal{A},V)}\right] _{G,Y\oplus \bar{\xi}}[v]_G=[\nabla_{W}v]_G$$
Observe that these so called quotient connections are not connections in the usual sense as the derivation is performed with respect to sections of $TQ/G$ instead of sections of $T(Q/G)$. Only the horizontal quotient connection can be thought as a usual connection via the horizontal lift. In addition one can check (see for example, \cite{LPcategory}) that
$$\left[ \nabla^{(\mathcal{A},V)}\right] _{G,Y\oplus \bar{\xi}}[v]_G=[\xi_v^V]_G,$$
where $\xi$ satisfies $\bar{\xi}=[\pi_{Q,V}(v),\xi]_G.$

\section{Lagrangian mechanics in the $\mathfrak{LP}$ category.} \label{LPsumary}
Lagrangian mechanics is generally set in $TQ$, the tangent bundle of the configuration space $Q$. Yet, the reduced Lagrangian obtained by Lagrange--Poincar\'e reduction of a $G$-invariant Lagrangian is defined on $TQ/G\cong T(Q/G)\oplus \tilde{\mathfrak{g}}$. To iterate this reduction procedure, a convenient category $\mathfrak{LP}$ of Lagrange--Poincar\'e bundles was introduced in \cite{CMR}, which includes $TQ/G$ and is stable under reduction.

The objects of $\mathfrak{LP}$ are vector bundles $TQ\oplus V\to Q$ obtained as a direct sum of the tangent bundle of a manifold $Q$ and a vector bundle $V\to Q$ on which there exist:
\begin{enumerate}[(i)]
\item \label{LP1a} a Lie bracket $[,]$ in the fibers of $V$;
\item \label{LP1b} a $V$-valued 2-form $\omega$ on $Q$;
\item \label{LP1c} a linear connection $\nabla$ on $V$;
\end{enumerate}
such that the bilineal operator defined by
$$[X_1\oplus w_1,X_2\oplus w_2]=[X_1,X_2]\oplus(\nabla_{X_1}w_2-\nabla_{X_2}w_1-\omega(X_1,X_2)+[w_1,w_2]),$$
satisfies the Jacobi indentity (that is, its is a Lie Bracket on sections $X\oplus w\in \Gamma(TQ\oplus V)$), where $[X_1,X_2]$ is the Lie bracket of vector fields and $[w_1,w_2]$ is the Lie bracket in the fibers of $V$. The morphisms of $\mathfrak{LP}$ are vector bundle morphisms preserving this extra structure as detailed in \cite{CMR}.
\begin{proposition} \label{quotientLP}\emph{\cite[\S 6.2]{CMR}}
Let $TQ\oplus V \to Q$ be an object of $\mathfrak{LP}$ with additional structure $[,]$, $\omega$ and $\nabla$. Let $\rho:G\times (TQ\oplus V)\to TQ\oplus V$ be a free and proper action in the category $\mathfrak{LP}$ (for all $g\in G$, $\rho_g$ is an isomorphism in $\mathfrak{LP}$) and $\mathcal{A}$ a principal connection on $Q\to Q/G$. Then, the vector bundle
$$T(Q/G)\oplus\tilde{\mathfrak{g}}\oplus(V/G)$$
with additional structures $[,]^{\tilde{\mathfrak{g}}}$, $\omega^{\tilde{\mathfrak{g}}}$ and $\nabla^{\tilde{\mathfrak{g}}}$ in $\tilde{\mathfrak{g}}\oplus(V/G)$ given by
\begin{align*}
\nabla^{\tilde{\mathfrak{g}}}_X(\bar{\xi}\oplus[v]_G)=&\nabla^{\mathcal{A}}_X\bar{\xi}\oplus\left([\nabla^{(\mathcal{A},H)}]_{G,X}[v]_G-[\omega]_G(X,\bar{\xi}) \right),\\
\omega^{\tilde{\mathfrak{g}}}(X_1,X_2)=&\tilde{B}(X_1,X_2)\oplus [\omega]_G(X_1,X_2), \\
\,[\bar{\xi}_1\oplus[v_1]_G,\bar{\xi}_2\oplus[v_2]_G]^{\tilde{\mathfrak{g}}} =& [\bar{\xi}_1,\bar{\xi}_2] \oplus \left([\nabla^{(\mathcal{A},V)}]_{G,\bar{\xi}_1} [v_2]_G\right. \\ &- \left. [\nabla^{(\mathcal{A},V)}]_{G,\bar{\xi}_2}[v_1]_G
-[\omega]_G(\bar{\xi}_1,\bar{\xi}_2)+[[v_1]_G,[v_2]_G]_G  \right)
\end{align*}
is an object of the $\mathfrak{LP}$ category called the reduced bundle with respect to the group $G$ and the connection $\mathcal{A}$.
\end{proposition}

Given a Lagrangian $L:TQ\oplus V\to\R$ defined on a Lagrange--Poincar\'e bundle, a curve $\dot{q}(t)\oplus v(t):[t_0,t_1]\to TQ\oplus V$ is said to be critical if and only if
$$0=\left.\frac{d}{d\varepsilon}\right|_{\varepsilon =0}\int_{t_0}^{t_1} L(\dot{q}_{\varepsilon}(t)\oplus v_{\varepsilon}(t)) dt,$$
where $\dot{q}_{\varepsilon}(t)\oplus v_{\varepsilon}(t)$ is a variation of $\dot{q}(t)\oplus v(t)$ such that $\delta\dot{q}$ is the lifted variation of a free variation $\delta q$ and
 $$\delta v=\frac{Dw}{dt}+[v,w]+\omega_q(\delta q,\dot{q}),$$
where $w(t)$ is a curve in $V$ with $w(t_0)=w(t_1)=0$ that projects to $q(t)$. This restricted variational principle is equivalent to the Lagrange--Poincar\'e equations
\begin{equation}
\frac{\delta L}{\delta q}-\frac{D}{Dt}\frac{\delta L}{\delta \dot{q}}=\left\langle\frac{\delta L}{\delta v},\omega_q(\dot{q},\cdot)\right\rangle ,
\end{equation}
\begin{equation}
\mathrm{ad}^*_{v}\frac{\delta L}{\delta v}=\frac{D}{Dt}\frac{\delta L}{\delta v},
\end{equation}
where for all $u,v\in V$ and $w\in V^*$, $\mathrm{ad}^*_{v}w(u)=w([v,u])$.

Suppose that a Lie group $G$ acts on ~$TQ\oplus V$ as in Proposition \ref{quotientLP} and that $L$ is $G$-invariant, so that it can be dropped to the quotient as a reduced Lagrangian $$l:T(Q/G)\oplus\tilde{\mathfrak{g}}\oplus(V/G)\to\R .$$
We denote by $\pi_G$ the projection of $TQ\oplus V \to (TQ\oplus V)/G$ and $\alpha^{TQ\oplus V}_\mathcal{A}$ the identification between $(TQ\oplus V)/G$ and $T(Q/G)\oplus\tilde{\mathfrak{g}}\oplus(V/G)$. As seen \cite{LPcategory}, a curve $\dot{q}(t)\oplus v(t)$ is critical for the variational problem set by $L$ if and only if the curve
$$\dot{x}(t)\oplus \bar{\xi}(t) \oplus [v]_G(t)=\alpha^{TQ\oplus V}_\mathcal{A}\circ \pi_G (\dot{q}(t)\oplus v(t)),$$
is critical for the (constrained) variational problem set by $l$. Equivalently, $\dot{q}(t)\oplus v(t)$ solves the Lagrange-Poincar\'e equations given by $L$ in $TQ\oplus V$ if and only if $\dot{x}(t)\oplus \bar{\xi}(t) \oplus [v]_G(t)$ solves the Lagrange--Poincar\'e equations given by $l$ in $T(Q/G)\oplus\tilde{\mathfrak{g}}\oplus(V/G)$. In other words, Lagrangian reduction can be performed on mechanical systems defined on Lagrange-Poincar\'e bundles. Furthermore, as the category is stable under reduction, the reduction process can be iterated. A $G$-invariant Lagrangian can be first reduced by normal subgroup $N$ of $G$, and afterwards by $K = G/N$. Whenever the connections implied are conveniently chosen, this chain of reductions by stages will result in an equivalent system to the one obtained directly reducing by $G$.

The Noether current of a Lagrangian system on a Lagrange-Poincar\'e bundle is $J:TQ\oplus V\to \mathfrak{g}^*$ such that
\begin{equation*}\label{NoetherCurrent}
J(\dot{q}\oplus v)(\eta)=\left\langle \frac{\partial L}{\partial \dot{q}}(\dot{q}\oplus v),\eta_q^Q\right\rangle ,
\end{equation*}
for any $\dot{q}\oplus v \in TQ \oplus V$ and any $\eta \in \mathfrak{g}$. In \cite{LPcategory} it is proved that its evolution along a solution of the Lagrange-Poincar\'e equations is given by
\begin{align}\label{Noetherdrift}
\frac{d}{dt}J(\dot{q}(t)\oplus v(t))(\eta)=-\left\langle\frac{\partial L}{\partial v}(\dot{q}(t)\oplus v(t)),\omega(\dot{q}(t),\eta_{q(t)}^Q)+\eta_{v(t)}^V\right\rangle.
\end{align}
Even if $L$ is $G$-invariant, the Noether current is not necessarily preserved. In fact, it can be proved that its drift is equivalent to the vertical Lagrange-Poincar\'e equation of the reduced system restricted to $\tilde{\mathfrak{g}}$.

\section{Rigid Body with Rotors. Euler-Poincar\'e Reduction}\label{rigidbody}
A rigid body with rotors is a mechanical system consisting of a rigid body that has moving pieces which are rotating around different axes. We will restrict ourselves to the case in which there are three rotors whose axes are the principal axes of inertia, although the results in this paper could be easily generalised to any number of rotors at different positions. The configuration bundle of such system is $Q=SO(3)\times \sS^1\times \sS^1 \times \sS^1$ encoding the spatial rotation of both the rigid body and the three rotors respectively. We denote an arbitrary point in $Q$ either $(R,\theta_1,\theta_2,\theta_3)$ or $(R,\theta)$.

\newcommand{\asa}{2}
\newcommand{\bsa}{0.5}
\newcommand{\csa}{0.5}
\tdplotsetmaincoords{70}{135}
%




The Lagrangian, $L$, for a free rigid body with rotors is a real function defined on $TQ$ as
\begin{equation*}\label{lag}
L=\frac{1}{2}\langle R^{-1}\dot{R}, IR^{-1}\dot{R}\rangle +\frac{1}{2}\langle R^{-1}\dot{R}+\dot\theta, K (R^{-1}\dot{R}+\dot\theta) \rangle
\end{equation*}
where $(R,\theta,\dot{R},\dot{\theta})\in TQ$, $I$ represents the inertia tensor of the rigid solid, $K$ is the inertia tensor of the rotors, and $\langle \cdot,\cdot\rangle$ represents the usual scalar product in $\R^3$. We have made the usual identification of the Lie algebras ($\so (3)$,$[,]$) and ($\R^3$,$\times$).

Since the configuration bundle of the rigid body with rotors is the Lie group $SO(3)\times \sS^1\times \sS^1 \times \sS^1$, there is an smooth action of the configuration bundle on itself by left translations which is free and proper. According to the theory of Euler-Poincar\'e reduction, the action lifts to $TQ$ and the lifted action is free and proper. In addition, the quotient of $TQ$ by this action is a smooth manifold identified with the Lie algebra via
\begin{equation*}
\begin{split}
T(SO(3)\times \sS^1\times \sS^1 \times \sS^1)/(SO(3)\times \sS^1\times \sS^1 \times \sS^1)\to &\so (3)\oplus \R^3 \\
[R,\theta,\dot{R},\dot{\theta}] \mapsto & (\Omega,\Omega_r)
\end{split}
\end{equation*}
where $\Omega=R^{-1}\dot{R}$ is the angular velocity of the rigid body and $\Omega_r=\dot\theta$ is the rotor angular velocity. As the free Lagrangian is invariant under the action on the configuration bundle, a reduced Lagrangian, $\ell:\so (3)\oplus \R^3 \to \R$, can then be written as
\begin{equation*}
\begin{split}
\ell(\Omega,\Omega_r)=&\frac{1}{2}\langle \Omega, I\Omega\rangle +\frac{1}{2}\langle \Omega+\Omega_r, K (\Omega+\Omega_r) \rangle\\ =&\frac{1}{2}\langle \Omega, (I+K)\Omega\rangle +\frac{1}{2}\langle \Omega_r, K\Omega_r \rangle+\langle \Omega, K\Omega_r \rangle
\end{split}
\end{equation*}
The evolution of the system is described by the reduced Euler-Poincar\'e equations
\begin{equation}\label{EP}
\frac{D}{dt}\left( \frac{\partial \ell}{\partial \Omega}\oplus \frac{\partial \ell}{\partial \Omega_r}\right)=\text{ad}^*_{(\Omega,\Omega_r)} \left( \frac{\partial \ell}{\partial \Omega}\oplus \frac{\partial \ell}{\partial \Omega_r}\right),
\end{equation}
where the coadjoint action refers to the Lie bracket such that for any two elements $(\alpha_1,\beta_1),(\alpha_2,\beta_2)$ in $\so (3)\oplus \R^3$;  $$[(\alpha_1,\beta_1),(\alpha_2,\beta_2)]=(\alpha_1\times\alpha_2,0).$$ Note that \eqref{EP} is defined in $(\so (3)\oplus \R^3)^*$ as
$\frac{\partial \ell}{\partial \Omega}=\langle (I+K)\Omega,\cdot\rangle+\langle K\Omega_r,\cdot\rangle$ and $\frac{\partial \ell}{\partial \Omega_r}=\langle K\Omega_r,\cdot\rangle + \langle K\Omega,\cdot\rangle$ both belong to that dual vector space. Then, equation \eqref{EP} applied to an element $(\alpha,\beta)\in\so (3)\oplus \R^3$ can be written as
$$\langle(I+K)\dot{\Omega},\alpha\rangle+\langle K \dot{\Omega}_r,\alpha \rangle+\langle K \dot{\Omega}_r,\beta\rangle+\langle K\dot{\Omega},\beta\rangle=\langle (I+K)\Omega + K\Omega_r,\Omega\times \alpha\rangle$$
As a consequence, the reduced equations of motion are
\begin{equation}\label{EP1}
(I+K)\dot{\Omega}+ K \dot{\Omega}_r=((I+K)\Omega + K\Omega_r)\times\Omega
\end{equation}
\begin{equation}\label{EP2}
 K \dot{\Omega}_r+ K\dot{\Omega}=0
\end{equation}
where \eqref{EP2} can be understood as the conservation of the rotors momentum.

\section{Reduction by stages initiated by $\sS^1\times \sS^1 \times \sS^1$} \label{stagess}
In this section we will reduce by stages the free rigid body with rotors first reducing by $\sS^1\times \sS^1 \times \sS^1$ and later on by $SO(3)$. In order to do the first step of reduction we choose a principal connection $A_{R,\theta} (\dot{R},\dot{\theta}) =R^{-1}\dot{R}+\dot{\theta}$ of the principal bundle $SO(3)\times\sS^1\times \sS^1 \times \sS^1\to SO(3)$ and we define the total angular velocity $\eta=\Omega+\Omega_r=A_{R,\theta} (\dot{R},\dot{\theta})$. This choice of connection coincides with the mechanical connection induced by the Riemannian metric on $SO(3)\times\sS^1\times \sS^1 \times \sS^1$ given by
$$\langle\langle (\dot{R},\dot{\theta}),(\dot{Q},\dot{\phi})\rangle\rangle_{(R,\theta)}=\langle R^{-1}\dot{Q},IR^{-1}\dot{R}\rangle+\langle R^{-1}\dot{Q}+\dot{\phi},K(R^{-1}\dot{R}+\dot{\theta})\rangle$$
and it is motivated by the simplification of the reduced Lagrangian that we get with it.
We apply this connection to the identification
\begin{align*}
\frac{T(SO(3)\times \sS^1\times \sS^1 \times \sS^1)}{\sS^1\times \sS^1 \times \sS^1}\to & T(SO(3))\oplus \tilde{\R}^3&\to & T(SO(3))\oplus (SO(3)\times\R^3)\\
[R,\theta,\dot{R},\dot{\theta}] \mapsto & (R,\dot{R},[R,\theta,\eta])&\mapsto & (R,\dot{R},\eta)
\end{align*}
where the right side is obtained by trivializing $\tilde{\R}^3$ as $SO(3)\times\R^3$. The bundle $T(SO(3))\oplus (SO(3)\times\R^3)$ is an object in the $\mathfrak{LP}$ category with trivial Lie bracket in the fibers of $SO(3)\times\R^3\to SO(3)$, trivial covariant derivative and a $\R^3$-valued 2-form on $SO(3)$, $\tilde{B}_R(\dot{R},\dot{Q})=(R^{-1}\dot{Q})\times(R^{-1}\dot{R})$ induced by
\begin{equation*}
\begin{split}
B_{R,\theta}((\dot{R},\dot{\theta}),(\dot{Q},\dot{\phi})&=-A_{R,\theta}([(\dot{R},\dot{\theta})^H,(\dot{Q},\dot{\phi})^H])\\&=-A_{R,\theta}([(\dot{R},-R^{-1}\dot{R}),(\dot{Q},-R^{-1}\dot{Q})])\\ &=-A_{R,\theta}(\dot{R}R^{-1}\dot{Q}-\dot{Q}R^{-1}\dot{R},0)=(R^{-1}\dot{Q})\times(R^{-1}\dot{R}).
\end{split}
\end{equation*}
The system obtained after the first step of reduction is defined in this $\mathfrak{LP}$-bundle and it is given by the reduced Lagrangian;
\begin{equation} \label{redlag}
l_{1,S}(R,\dot{R},\eta)= \frac{1}{2} \langle R^{-1}\dot{R}, IR^{-1}\dot{R}\rangle+\frac{1}{2}\langle \eta, K\eta \rangle.
\end{equation}
On one hand, since the group $\sS^1\times \sS^1 \times \sS^1$ is abelian, $\frac{\partial l_{1,S}}{\partial \eta}=\langle K\eta,\cdot\rangle$ and $\frac{D}{dt}\left( \frac{\partial l_{1,S}}{\partial \eta}\right)=\langle K\dot{\eta},\cdot\rangle$, the vertical Lagrange-Poincar\'e equation $$\frac{D}{dt}\left( \frac{\partial l_{1,S}}{\partial \eta}\right) -\text{ad}^*_{\bar{\eta}}\left(\frac{\partial l_{1,S}}{\partial \eta}\right)=0$$ can be written as
\begin{equation}\label{redsver}
K\dot{\eta}=0,
\end{equation}
which, as we already mentioned, can be interpreted as the conservation of the momentum of the rotors. On the other hand, since $\langle\frac{\partial l_{1,S}}{\partial\eta},\tilde{B}_R(\dot{R},\dot{Q})\rangle=\langle K\eta, (R^{-1}\dot{Q})\times (R^{-1}\dot{R})\rangle=\langle R^{-1}\dot{Q},(R^{-1}\dot{R})\times K\eta\rangle$ the horizontal Lagrange-Poincar\'e equation $$\frac{D}{dt}\left( \frac{\partial l_{1,S}}{\partial \dot{R}}\right)-\frac{\partial l_{1,S}}{\partial R}+\left\langle \frac{\partial l_{1,S}}{\partial\eta},\tilde{B}_R(\dot{R},\cdot)\right\rangle=0$$  is written as
\begin{equation}\label{redshor}
I\dot{\Omega}-I\Omega\times\Omega-(K\eta)\times\Omega=0
\end{equation}
which is the equation of a rigid body with inertia tensor $I$ with an additional term $(K\eta)\times\Omega$ coming from the curvature of the connection used to reduce.

The second step of reduction uses the action of $SO(3)$ in $T(SO(3))\oplus (SO(3)\times\R^3)$ given by $\Gamma\cdot(R,\dot{R},\eta)=(\Gamma R,\Gamma\dot{R},\eta)$ together with the connection $A^{SO(3)}_R(\dot{R})=R^{-1}\dot{R}=\Omega$. This gives the isomorphism
\begin{align*}
T(SO(3))\oplus (SO(3)\times\R^3)/SO(3) \to &\mathfrak{so}(3)\times\R^3\\
[R,\dot{R},\eta] \mapsto &(\Omega,\eta).
\end{align*}
As the base space resulting from this step of reduction is a single point, the induced connection $\nabla^{\tilde{\so}(3)}=0$ and the 2-form  $\omega^{\tilde{\so}(3)}=0$. It can be seen as well that $[\nabla^{(A,H)}]_{SO(3)}$ is trivial and consequently the action of $SO(3)$ in $T(SO(3))\oplus (SO(3)\times\R^3)$ is horizontal in the sense that $[\nabla^{(A,V)}]_{SO(3),\Gamma}[\eta]=[\Gamma^V_{\eta}]=0$ for all $\Gamma\in SO(3)$. Therefore, given $\Omega_1,\Omega_2 \in \so (3)$ and $\eta_1,\eta_2 \in \R^3$ the induced bracket is
\begin{align*}
[(\Omega_1,\eta_1),(\Omega_2,\eta_2)]^{\tilde{\so}(3)}=&[\Omega_1,\Omega_2]\oplus ([\nabla^{(A^{SO(3)},V)}]_{SO(3),\Omega_1}[\eta_2]\\
&-[\nabla^{(A^{SO(3)},V)}]_{SO(3),\Omega_2}[\eta_1]-[\tilde{B}]_R(\Omega_1,\Omega_2)+[\eta_1,\eta_2])\\
=&(\Omega_1\times\Omega_2)\oplus(\Omega_1\times\Omega_2)
\end{align*}
Thus, the second reduced space is the $\mathfrak{LP}$-bundle over a point $\mathfrak{so}(3)\times\R^3\to \{\bullet\}$ with $\nabla^{\tilde{\so}(3)}=0$, $\omega^{\tilde{\so}(3)}=0$ and $[(\Omega_1,\eta_1),(\Omega_2,\eta_2)]^{\tilde{\so}(3)}= (\Omega_1\times\Omega_2)\oplus(\Omega_1\times\Omega_2)$. In other words, the reduced configuration space is the Lie algebra $\mathfrak{so}(3)\times\R^3$ with bracket given by $[,]^{\tilde{\so}(3)}$. In addition, the reduced Lagrangian is $$l_{2,S}(\Omega,\eta)= \frac{1}{2} \langle \Omega, I\Omega\rangle+\frac{1}{2}\langle \eta, K\eta \rangle.$$

There is no horizontal Lagrange-Poincar\'e equation, and with respect to the vertical we have $$\frac{D}{dt}\left( \frac{\partial l_{2,S}}{\partial \Omega}\oplus \frac{\partial l_{2,S}}{\partial \eta}\right)=\text{ad}^*_{\Omega,\eta}\left(\frac{\partial l_{2,S}}{\partial \Omega}\oplus \frac{\partial l_{2,S}}{\partial \eta}\right)=0.$$ Applying this equation to an arbitrary element $(\alpha,\beta)\in \mathfrak{so}(3)\times\R^3$ and using that $\frac{\partial l_{2,S}}{\partial \Omega}=\langle I\Omega,\cdot\rangle$, $\frac{\partial l_{2,S}}{\partial \eta}=\langle K\eta,\cdot\rangle$, the vertical equation can be written as
$$\langle I\dot{\Omega},\alpha\rangle +\langle K\dot{\eta},\beta\rangle=\langle I \Omega,\Omega\times\alpha\rangle+\langle K\eta,\Omega\times\alpha\rangle$$
from where we obtain the following motion equations
\begin{equation}\label{redssover1}
I\dot{\Omega}=I\Omega\times\Omega+(K\eta)\times\Omega
\end{equation}
and
\begin{equation}\label{redssover2}
K\dot{\eta}=0.
\end{equation}
Vertical equations from the second step of reduction (equations \eqref{redssover1} and \eqref{redssover2}) coincide with the horizontal and vertical ones of the first step of reduction (equations \eqref{redsver} and \eqref{redshor}) showing that the systems obtained in both steps of reduction are equivalent. Notice that equation \eqref{redshor} is an horizontal equation in the first step of reduction whereas in the second step it is obtained as the vertical equation \eqref{redssover1}. These descriptions are in turn equivalent to the set of equations obtained by Euler-Poincar\'e reduction in Section \ref{EP}. Indeed, equation \eqref{EP2} coincides with equation \eqref{redssover2} and equation \eqref{EP1} is the sum of \eqref{redssover1} and \eqref{redssover2}.

The $\sS^1\times\sS^1\times\sS^1$-invariance of the Lagrangian $L$ in \ref{lag} is used in the first step of reduction and the associated Noether current is:
\begin{align*}
J_1:T(SO(3)\times\sS^1\times \sS^1 \times \sS^1 ) \to &(\R^3)^*\\
(R,\theta,\dot{R},\dot{\theta})\mapsto &(a\mapsto \langle \frac{\partial L}{\partial\dot{R} }\oplus\frac{\partial L}{\partial\dot{\theta}} ,a_{(R,\theta)}^{\sS^1\times \sS^1 \times \sS^1\times SO(3)}\rangle )
\end{align*}
Identifying $\R^3$ with its dual and taking into account that $\frac{\partial L}{\partial\dot{\theta}}=K(R^{-1}\dot{R}+\dot{\theta})$ and $a_{(R,\theta)}^{\sS^1\times \sS^1 \times \sS^1\times SO(3)}=(R,\theta,0,a)$,
\begin{align*}
J_1:T(SO(3)\times\sS^1\times \sS^1 \times \sS^1 ) \to &\R^3\\
(R,\theta,\dot{R},\dot{\theta})\mapsto &K(R^{-1}\dot{R}+\dot{\theta}).
\end{align*}
This Noether current can be interpreted as the momentum of the rotors, it is conserved and its conservation is equivalent to equation \eqref{redsver}. The $SO(3)$-invariance of the reduced Lagrangian $l_{1,S}$ in \eqref{redlag} is used in the second step of reduction to which we associate the Noether current
\begin{align*}
J_2:T(SO(3))\oplus (SO(3)\times\R^3) \to &\so(3)^*\\
(R,\dot{R},\eta)\mapsto &(b\mapsto \langle \frac{\partial l_{1,S}}{\partial\dot{R} },b_{R}^{SO(3)}\rangle).
\end{align*}
Since $b_{R}^{SO(3)}=(R,bR)$, the identification of $\so(3)$ with its dual gives
\begin{align*}
J_2:T(SO(3))\oplus (SO(3)\times\R^3) \to &\so(3)\\
(R,\dot{R},\eta)\mapsto & RIR^{-1}\dot{R}R^{-1}.
\end{align*}
This Noether current is interpreted as the angular momentum of a rigid body with inertia tensor $I$. Since there are rotors, this is not the total angular momentum of the system and it is not preserved. Particularization of the drift of the Noether current in equation \eqref{Noetherdrift} shows that for every $b\in\so(3)$
\begin{equation}
\frac{d}{dt}\langle J(R,\dot{R},\eta),b\rangle=-\langle K \eta, (R^{-1}bR)\times R^{-1}R\rangle
\end{equation}
which can be seen to be equivalent to equation \eqref{redssover1} which is the vertical equation of the second step of reduction that was obtained as horizontal equation \eqref{redshor} in the first reduction.

\section{Reduction by stages initiated by $SO(3)$} \label{stagesso}
In this section we will reduce by stages starting by the reduction with respect to $SO(3)$ and later by $\sS^1\times \sS^1 \times \sS^1$. This will be done in two different ways. In the first method we shall use in each step the Maurer-Cartan connection as in \cite{hindi}, while in the second method a mechanical connection like the one presented in \cite{MScheurle} will be used.

\subsection{Reduction with Maurer-Cartan connection}
We consider the natural lift to $T(\sS^1\times \sS^1 \times \sS^1\times SO(3))$ of the action of $SO(3)$ on itself on the second factor of $\sS^1\times \sS^1 \times \sS^1\times SO(3)$. The principal bundle $\sS^1\times \sS^1 \times \sS^1\times SO(3)\to\sS^1\times \sS^1 \times \sS^1$ has a principal connection $A_{R,\theta}(\dot{R},\dot{\theta})=\text{Ad}_{R}(R^{-1}\dot{R})=\dot{R}R^{-1}$, that is, the Maurer-Cartan form. We then have
\begin{align*}
T(SO(3)\times \sS^1\times \sS^1 \times \sS^1)/SO(3)\to &T(\sS^1\times \sS^1 \times \sS^1)\oplus \tilde{\so} (3) \\
[R,\theta,\dot{R},\dot{\theta}] \mapsto & (\theta,\dot{\theta},[R,\theta,\dot{R}R^{-1}])
\end{align*}
which in turn can be trivialized to
\begin{align*}
T(\sS^1\times \sS^1 \times \sS^1)\oplus \tilde{\so} (3) \to &T(\sS^1\times \sS^1 \times \sS^1)\oplus (\sS^1\times \sS^1 \times \sS^1\times\so (3))\\
(\theta,\dot{\theta},[R,\theta,\dot{R}R^{-1}])\mapsto &(\theta,\dot{\theta},\Omega=R^{-1}\dot{R}),
\end{align*}
where $\Omega=R^{-1}R$ as in section \ref{rigidbody}. The reduced space $T(\sS^1\times \sS^1 \times \sS^1)\oplus \tilde{\so} (3)$ is an $\mathfrak{LP}$-bundle with covariant derivative in $\tilde{\so} (3)$ given by $$\frac{D}{dt}[R(t),\theta(t),\alpha(t)]=[R(t),\theta(t),\dot{\alpha}(t)-\Omega\times\alpha(t)],$$ Lie bracket in $\tilde{\so} (3)$ induced by the usual cross product, and, since the Maurer-Cartan connection is integrable, the $\tilde{\so} (3)$-valued 2-form on $\sS^1\times \sS^1 \times \sS^1$ is zero. The trivialization  $T(\sS^1\times \sS^1 \times \sS^1)\oplus (\sS^1\times \sS^1 \times \sS^1\times\so (3))$ is also an $\mathfrak{LP}$-bundle with null $(\sS^1\times \sS^1 \times \sS^1\times\so (3))$-valued 2-form, Lie bracket in $\sS^1\times \sS^1 \times \sS^1\times\so (3)$ induced by the usual cross product, yet, the covariant derivative can be easily seen to be trivial.

The reduced Lagrangian is
\begin{equation}\label{redlagMC}
l_{1,O}(\theta,\dot{\theta},\Omega)= \frac{1}{2} \langle\Omega, (I+K)\Omega\rangle+\frac{1}{2}\langle \dot{\theta}, K\dot{\theta} \rangle+\frac{1}{2}\langle \Omega, K\dot{\theta} \rangle,
\end{equation}
which has a non-decoupled term depending on the rotor and the rigid body. This does not alter the fact that Lagrange-Poincar\'e equations can still be obtained.

The vertical equation $\frac{D}{dt}\left( \frac{\partial l_{1,O}}{\partial \Omega}\right) -\text{ad}^*_{\Omega}\left(\frac{\partial l_{1,O}}{\partial \Omega}\right)=0$ can be rewritten as $$\langle (I+K)\dot{\Omega}+K\ddot{\theta},\alpha\rangle=\langle(I+K)\Omega+K\dot{\theta},\Omega\times\alpha\rangle,$$ where $\alpha$ ia an arbitrary element of $\so (3)$. That is, we have
\begin{equation}\label{redsover1}
(I+K)\dot{\Omega}+K\ddot{\theta}=((I+K)\Omega+K\dot{\theta})\times\Omega.
\end{equation}
With respect to the horizontal equation $$\frac{D}{dt}\left( \frac{\partial l_{1,O}}{\partial \dot{\theta}}\right)-\frac{\partial l_{1,O}}{\partial \theta}+\langle\frac{\partial l_{1,O}}{\partial\Omega},\tilde{B}_{\theta}(\dot{\theta},\cdot)\rangle=0,$$ since $l_{1,O}$ does not depend on $\theta$ and $\frac{\partial l_{1,O}}{\partial \dot{\theta}}=\langle K\dot{\theta}+K\Omega,\cdot\rangle$, it reads
\begin{equation}\label{redsohor1}
K\ddot{\theta}+K\dot{\Omega}=0.
\end{equation}

The equations of motion \eqref{redsover1} and \eqref{redsohor1} obtained in the first step of reduction are, respectively, equivalent to equations \eqref{EP1} and \eqref{EP2}, obtained in the Euler-Poincar\'e reduction. We now perform the second step of reduction. The group $\sS^1\times \sS^1 \times \sS^1$ acts on $T(\sS^1\times \sS^1 \times \sS^1)\oplus (\sS^1\times \sS^1 \times \sS^1\times\so (3))$ as the product of the lift action in the first factor and the natural one in the second, that is, given $\varphi\in \sS^1\times \sS^1 \times \sS^1$, $\varphi\cdot(\theta,\dot{\theta},\Omega)=(\theta+\varphi,\dot{\theta},\Omega)$. The connection used to reduce is $A_{\theta}(\dot{\theta})=\dot{\theta}=\Omega_r$ and induces an isomorphism
\begin{align*}
T(\sS^1\times \sS^1 \times \sS^1)\oplus (\sS^1\times \sS^1 \times \sS^1\times\so (3))/(\sS^1\times \sS^1 \times \sS^1)\to& \R^3\oplus\so (3)\\
[\theta,\dot{\theta},\Omega] \mapsto& (\Omega_r=\dot{\theta},\Omega).
\end{align*}
This time, the second reduced space is the $\mathfrak{LP}$-bundle $\R^3\times \mathfrak{so}(3)\to\{\bullet\}$ over a point with $\nabla^{\tilde{\R}^3}=0$ and $\omega^{\tilde{\R}^3}=0$. However, in contrast to the second step of reduction in Section \ref{stagess}, $[(\Omega_{r,1},\Omega_1),(\Omega_{r,2},\Omega_2)]^{\tilde{\R}^3}= 0\oplus(\Omega_1\times\Omega_2)$ and the reduced Lagrangian is $$l_{2,O}(\Omega_r,\Omega)= \frac{1}{2} \langle \Omega, (I+K)\Omega\rangle+\frac{1}{2}\langle \Omega_r, K\Omega_r \rangle+\frac{1}{2}\langle \Omega, K\Omega_r \rangle.$$

There are no horizontal equation, and the vertical equations  $$\frac{D}{dt}\left( \frac{\partial l_{2,O}}{\partial \Omega_r}\oplus \frac{\partial l_{2,O}}{\partial \Omega}\right)=\text{ad}^*_{\Omega_r,\Omega}\left(\frac{\partial l_{2,O}}{\partial \Omega_r}\oplus \frac{\partial l_{2,O}}{\partial \Omega}\right)=0$$ are
\begin{equation}\label{redsosver}
(I+K)\dot{\Omega}+K\dot{\Omega_r}=((I+K)\Omega+K\Omega_r)\times\Omega.
\end{equation}
\begin{equation}\label{redsosvera}
K\dot{\Omega_r}+K\dot{\Omega}=0
\end{equation}
which are the same as the Lagrange-Poincar\'e equations of the first step of reduction labeled as \eqref{redsover1} and \eqref{redsohor1} once the change of variable $\dot{\theta}=\Omega_r$ is done. As in Section \ref{stagess}, the horizontal equation of the first step of reduction is obtained as a vertical equation in the second step of reduction.

\subsection{Reduction with the mechanical connection}
We continue working with the same action of $SO(3)$ on $T(\sS^1\times\sS^1\times\sS^1\times SO(3))$ as in the previous case. However, we shall use the connection $A_{R,\theta}(\dot{R},\dot{\theta})=\text{Ad}_{R}(R^{-1}\dot{R}+(I+K)^{-1}K\dot{\theta})$ which is the mechanical connection induced on the principal bundle $\sS^1\times \sS^1 \times \sS^1\times SO(3)\to\sS^1\times \sS^1 \times \sS^1$ by the metric on $\sS^1\times \sS^1 \times \sS^1\times SO(3)$ that makes $$L(R,\dot{R},\theta,\dot{\theta})= \frac{1}{2} \langle R^{-1}\dot{R}, IR^{-1}\dot{R}\rangle+\frac{1}{2}\langle R^{-1}\dot{R}+\dot{\theta}, K(R^{-1}\dot{R}+\dot{\theta}) \rangle$$ a kinetic energy term. The isomorphism induced by this connection is
\begin{align*}
T(SO(3)\times\sS^1\times\sS^1\times\sS^1)/SO(3)\to &T(\sS^1\times\sS^1\times\sS^1)\oplus\tilde{\so}(3) \\
[R,\theta,\dot{R},\dot{\theta}] \mapsto & (\theta,\dot{\theta},[R,\theta,\dot{R}R^{-1}+R(I+K)^{-1}K\dot{\theta}R^{-1}])
\end{align*}
which can then be trivialized to
\begin{align*}
T(\sS^1\times\sS^1\times\sS^1)\oplus\tilde{\so}(3)\to &T(\sS^1\times\sS^1\times\sS^1)\oplus(\sS^1\times \sS^1\times\sS^1\times\so(3))\\
(\theta,\dot{\theta},[R,\theta,A_{R,\theta}(\dot{R},\dot{\theta})])\mapsto &(\theta,\dot{\theta},\xi=\text{Ad}_{R^{-1}}(A_{R,\theta}(\dot{R},\dot{\theta}))),
\end{align*}
where a new variable $\xi=R^{-1}\dot{R}+(I+K)^{-1}K\dot{\theta}\in\so (3)$ has been introduced.

The reduced space $T(\sS^1\times \sS^1 \times \sS^1)\oplus \tilde{\so} (3)$ is an $\mathfrak{LP}$-bundle with covariant derivative in $\tilde{\so} (3)$ given by $$\frac{D}{dt}[R(t),\theta(t),\alpha(t)]=[R(t),\theta(t),\dot{\alpha}(t)-(\dot{R}R^{-1}+R(I+K)^{-1}K\dot{\theta}R^{-1})\times\alpha(t)]$$ and Lie bracket in $\tilde{\so} (3)$ induced by the usual cross product. The trivialization  $T(\sS^1\times \sS^1 \times \sS^1)\oplus (\sS^1\times \sS^1 \times \sS^1\times\so (3))$ is also an $\mathfrak{LP}$-bundle with covariant derivative in $\tilde{\so} (3)$ given by $$\frac{D}{dt}[R(t),\theta(t),\alpha(t)]=[R(t),\theta(t),\dot{\alpha}(t)-((I+K)^{-1}K\dot{\theta})\times\alpha(t)]$$ and the same Lie bracket. The $(\sS^1\times \sS^1 \times \sS^1\times\so (3))$-valued 2-form associated to the trivialization comes from the curvature of the mechanical connection and is given by
$$\tilde{B}_{\theta}(\dot{\theta},\dot{\varphi})=((I+K)K\dot{\varphi})\times((I+K)K\dot{\theta}).$$ Additionally, the reduced Lagrangian on $T(\sS^1\times \sS^1 \times \sS^1)\oplus (\sS^1\times \sS^1 \times \sS^1\times\so (3))$ is
\begin{equation}\label{redlagMECH}
l_{1,O}(\theta,\dot{\theta},\xi)= \frac{1}{2} \langle \xi, (I+K)\xi\rangle+\frac{1}{2}\langle K\dot{\theta}, (I+K)^{-1}I\dot{\theta} \rangle.
\end{equation}
Observe that, in contrast to the approach with the Maurer-Cartan connection in the previous section, this Lagrangian is decoupled in the sense that there is no term depending on both coordinates of the rigid body and the rotors. This is the main reason why the mechanical connection is used in \cite{MScheurle}, yet, the previous example shows that equations of motion can be written even for a coupled Lagrangian.

We now write explicitly the vertical Lagrange-Poincar\'e equation $$\frac{D}{dt}\left( \frac{\partial l_{1,O}}{\partial \bar{\xi}}\right) -\text{ad}^*_{\bar{\xi}}\left(\frac{\partial l_{1,O}}{\partial \bar{\xi}}\right)=0.$$ Let $\alpha\in\so(3)$, then
\begin{align*}
\langle&\frac{D}{dt}\left(\frac{\partial l_{1,O}}{\partial \xi} \right),\alpha\rangle=\frac{d}{dt}\langle\left(\frac{\partial l_{1,O}}{\partial \xi} \right),\alpha\rangle-\langle\frac{\partial l_{1,O}}{\partial \xi} ,\frac{D\alpha}{dt}\rangle=\\
=&\langle (I+K)\dot{\xi},\alpha\rangle+\langle(I+K)\xi,\dot{\alpha}\rangle-\langle(I+K)\xi,\dot{\alpha}\rangle+\langle(I+K)\xi,((I+K)^{-1}K\dot{\theta})\times\alpha\rangle\\
=&\langle (I+K)\dot{\xi},\alpha\rangle+\langle(I+K)\xi,((I+K)^{-1}K\dot{\theta})\times\alpha\rangle
\end{align*}
and
\begin{align*}
\langle\text{ad}^*_{\bar{\xi}}\left(\frac{\partial l_{1,O}}{\partial \bar{\xi}}\right),\alpha\rangle=\langle(I+K)\xi,\xi\times\alpha\rangle=\langle((I+K)\xi)\times\xi,\alpha\rangle.
\end{align*}
Hence, the vertical Lagrange-Poincar\'e equation is
\begin{equation}\label{redsover2}
(I+K)\dot{\xi}=((I+K)\xi)\times(\xi-(I+K)^{-1}K\dot{\theta})=((I+K)\xi)\times\Omega.
\end{equation}

The explicit expression of the horizontal Lagrange-Poincar\'e equations,
\begin{equation*}
\frac{D}{dt}\left( \frac{\partial l_{1,O}}{\partial \dot{\theta}}\right)-\frac{\partial l_{1,O}}{\partial \theta}+\langle\frac{\partial l_{1,O}}{\partial\xi},\tilde{B}_{\theta}(\dot{\theta},\cdot)\rangle=0,
\end{equation*}
is more involved since the term $\frac{\partial l_{1,O}}{\partial \theta}$ is distinct from zero in spite that the Lagrangian does not depend on $\theta$. This is because $\frac{\partial l_{1,O}}{\partial \theta}$ is not a simple partial derivative but a derivative with respect to a horizontal lift (for example, see \cite[\S 3]{CMR} or \cite[\S 3]{LPcategory}). We shall denote temporarily $\frac{\partial^c l_{1,O}}{\partial \theta}$ this derivative to distinguish it from the partial derivative. The derivative at point $(\theta,\dot{\theta},\xi)$ applied to $(\varphi,\dot{\varphi})$ is
$$\frac{\partial^c l_{1,O}}{\partial \theta}(\theta,\dot{\theta},\xi)(\varphi,\dot{\varphi})=\frac{d}{ds}_{s=0}l_{1,O}(\varphi(s),u(s),v(s)),$$
where $\varphi(s)$ is a curve in $\sS^1\times \sS^1 \times \sS^1$ such that $\varphi(0)=\varphi$, $\dot{\varphi}(0)= \dot{\varphi}$, $u(s)$ is the horizontal lift of $\varphi(s)$ to $T(\sS^1\times \sS^1 \times \sS^1)$ through $(\theta,\dot{\theta})$ and  $v(s)$ is the horizontal lift of $\varphi(s)$ to $\sS^1\times \sS^1 \times \sS^1\times \so (3)$ through $(\theta,\xi)$. Then,
\begin{align*}
 \frac{\partial^c l_{1,O}}{\partial \theta}(\theta,\dot{\theta},\xi)(\varphi,\dot{\varphi})&=\left.\frac{d}{ds}\right\vert_{s=0}l_{1,O}(\varphi(s),u(s),v(s))\\
  &=\langle\frac{\partial l_{1,O}}{\partial \theta},\dot{\varphi}(s)\rangle+\langle\frac{\partial l_{1,O}}{\partial \dot{\theta}},\dot{u}(s)\rangle+\langle\frac{\partial l_{1,O}}{\partial \xi},\dot{v}(s)\rangle\\
  &=\langle\frac{\partial l}{\partial \xi},((I+K)^{-1}K\dot{\varphi}(0))\times v(0)\rangle\\
  &=\langle(I+k)\xi,((I+K)^{-1}K\dot{\varphi})\times\xi\rangle
\end{align*}
where we have used that the partial derivative $\frac{\partial l}{\partial \theta}=0$, $u(s)=\dot{\theta}$ and $\dot{v}(s)=((I+K)^{-1}K\dot{\varphi}(s))\times v(s)$. Denoting again the connection-dependent derivative as in \cite{LPcategory} and \cite{CMR},
$$\frac{\partial l_{1,O}}{\partial \theta}(\dot{\varphi})=\langle\xi\times((I+k)\xi),(I+K)^{-1}K\dot{\varphi}\rangle$$
Since
$$\frac{D}{dt}\left( \frac{\partial l_{1,O}}{\partial \dot{\theta}}\right)=\langle K(I+K)^{-1}I\ddot{\theta},\cdot\rangle$$
and
$$\langle \frac{\partial l_{1,O}}{\partial \xi},\tilde{B}_{\theta}(\dot{\theta},\cdot)\rangle=\langle (I+K)\xi,((I+K)^{-1}K\cdot)\times ((I+K)^{-1}K\dot{\theta})\rangle$$
The vertical Lagrange-Poincar\'e equations applied to $\dot{\varphi}$ can be written as
\begin{align*}
0=&\langle K(I+K)^{-1}I\ddot{\theta},\dot{\varphi}\rangle-\langle\xi\times((I+k)\xi),(I+K)^{-1}K\dot{\varphi}\rangle\\
&+\langle (I+K)\xi,((I+K)^{-1}K\dot{\varphi})\times ((I+K)^{-1}K\dot{\theta})\rangle \\
=&\langle I\ddot{\theta},(I+K)^{-1}K\dot{\varphi}\rangle-\langle\xi\times((I+k)\xi),(I+K)^{-1}K\dot{\varphi}\rangle\\
&+\langle((I+K)^{-1}K\dot{\varphi}),((I+K)^{-1}K\dot{\theta})\times ((I+K)\xi)\rangle \\
=&\langle I\ddot{\theta}-\xi\times((I+K)\xi)+((I+K)^{-1}K\dot{\theta})\times ((I+K)\xi),(I+K)^{-1}K\dot{\varphi})
\end{align*}
Consequently,
\begin{equation}\label{redsohor2}
I\ddot{\theta}=-((I+K)\xi)\times\Omega.
\end{equation}
The equations obtained in this first step of reduction with the mechanical connection are equivalent to the ones obtained with Euler-Poincar\'e reduction. In fact, equation \eqref{EP1} is the same as equation \eqref{redsover2} writen in terms of the new variable $\xi$ and equation \eqref{redsohor2} is the difference between equations \eqref{EP1} and \eqref{EP2}.

We finally proceed to the second step of reduction.  For any $\varphi\in \sS^1\times \sS^1 \times \sS^1$, the action of group $\sS^1\times \sS^1 \times \sS^1$  on $T(\sS^1\times \sS^1 \times \sS^1)\oplus (\sS^1\times \sS^1 \times \sS^1\times\so (3))$ is given by $\varphi\cdot(\theta,\dot{\theta},\Omega)=(\theta+\varphi,\dot{\theta},\Omega)$ and the connection used to reduce is simply $A_{\theta}(\dot{\theta})=\dot{\theta}=\Omega_r$. This induces an isomorphism
\begin{align*}
T(\sS^1\times \sS^1 \times \sS^1)\oplus (\sS^1\times \sS^1 \times \sS^1\times\so (3))/(\sS^1\times \sS^1 \times \sS^1)\to& \R^3\oplus\so (3)\\
[\theta,\dot{\theta},\xi] \mapsto& (\Omega_r=\dot{\theta},\xi).
\end{align*}
The second reduced space is the $\mathfrak{LP}$-bundle $\mathfrak{so}(3)\times\R^3\to \{\bullet\}$ over a point with $\nabla^{\tilde{\R}^3}=0$ and $\omega^{\tilde{\R}^3}=0$. The covariant derivative in $\sS^1\times \sS^1 \times \sS^1\times\so (3)$ reduces to an horizontal quotient connection, $[\nabla^{(A,H)}]_{\sS^1\times \sS^1 \times \sS^1}\xi=\dot{\xi}$, and a vertical quotient connection, $[\nabla^{(A,V)}]_{\sS^1\times \sS^1 \times \sS^1,\Omega_r}\xi=-((I+K)^{-1}K\Omega_r)\times\xi$. Hence,
\begin{align*}
[(\Omega_{r,1},\xi_1),(\Omega_{r,2},\xi_2)]^{\tilde{\R}^3}=& [\Omega_{r,1},\Omega_{r,2}]\oplus([\nabla^{(A,V)}]_{\sS^1\times \sS^1 \times \sS^1,\Omega_{r,1}}\xi_2\\&-[\nabla^{(A,V)}]_{\sS^1\times \sS^1 \times \sS^1,\Omega_{r,2}}\xi_1-[\tilde{B}]_{\theta}(\Omega_{r,1},\Omega_{r,2})+[\xi_1,\xi_2])\\
=&0\oplus -((I+K)^{-1}K\Omega_r)\times\xi_2+((I+K)^{-1}K\Omega_{r,2})\times\xi_1\\&+((I+K)^{-1}K\Omega_{r,2})\times((I+K)^{-1}K\Omega_{r,1})+\xi_1\times\xi_2.
\end{align*}
The reduced Lagrangian in $\mathfrak{so}(3)\times\R^3$ is $$l_{2,O}(\Omega_r,\xi)= \frac{1}{2} \langle \xi, (I+K)\xi\rangle+\frac{1}{2}\langle K\Omega_r, (I+K)^{-1}I\Omega_r \rangle$$
There are clearly no horizontal Lagrange-Poincar\'e equations, and the vertical equations $$\frac{D}{dt}\left( \frac{\partial l_{2,O}}{\partial \Omega_r}\oplus \frac{\partial l_{2,O}}{\partial \xi}\right)=\text{ad}^*_{\Omega_r,\xi}\left(\frac{\partial l_{2,O}}{\partial \Omega_r}\oplus \frac{\partial l_{2,O}}{\partial \Omega}\right)=0$$
can be written more explicitly as
\begin{equation}\label{redsosver2}
I\dot{\Omega}_r=-((I+K)\xi)\times\Omega,
\end{equation}
\begin{equation}\label{redsosvera2}
(I+K)\dot{\xi}=((I+K)\xi)\times\Omega,
\end{equation}
which are the same as the Lagrange-Poincar\'e equations obtained in the first step of reduction after the change of variable $\dot{\theta}=\Omega_r$.

\subsection{Noether currents}
When reduction by stages of the Lagrangian $L$ in \eqref{lag} begins with $SO(3)$-invariance, irrespective of the connection used, the associated Noether current is
\begin{align*}
J_1:T(SO(3)\times\sS^1\times \sS^1 \times \sS^1) \to &\so(3)^*\\
(R,\theta,\dot{R},\dot{\theta})\mapsto &(b\mapsto \langle \frac{\partial L}{\partial\dot{R} }\oplus\frac{\partial L}{\partial\dot{\theta}} ,b_{(R,\theta)}^{\sS^1\times \sS^1 \times \sS^1\times SO(3)}\rangle ).
\end{align*}
Since $\frac{\partial L}{\partial R}=R(I+K)R^{-1}\dot{R}+RK\dot{\theta})$ and $b_{(R,\theta)}^{\sS^1\times \sS^1 \times \sS^1\times SO(3)}=(R,\theta,bR,0)$, identification of $\so(3)$ with its dual gives
\begin{align*}
J_1:T(T(SO(3)\times\sS^1\times \sS^1 \times \sS^1) \to &\so(3)\\
(R,\theta,\dot{R},\dot{\theta})\mapsto &R(I+K)R^{-1}\dot{R}R^{-1}+RK\dot{\theta}R^{-1}=R\xi R^{-1}.
\end{align*}
This Noether current can be interpreted as the total momentum of the rigid body with rotors and its conservation is equivalent to equations \eqref{redsover1} and \eqref{redsover2}.

In contrast, in the second step of reduction the Noether current depends on the connection used in the first step of reduction. In this example, if the Maurer-Cartan connection is used in the first step, the Noether current obtained from the Lagragian \eqref{redlagMC} is
\begin{align*}
J_{2,MC}:T(\sS^1\times \sS^1 \times \sS^1)\oplus (\sS^1\times \sS^1 \times \sS^1\times\so(3)) \to &\R^3\\
(\theta,\dot{\theta},\Omega)\mapsto &K\dot{\theta}+K\Omega,
\end{align*}
after identification of $\R^3$ with its dual. This Noether current is the momentum of the rotors. As the Maurer-Cartan connection is flat and the action of $\sS^1\times \sS^1 \times \sS^1$ is horizontal on $\sS^1\times \sS^1 \times \sS^1\times\so(3)\to\sS^1\times \sS^1 \times \sS^1$, the second term in equation \eqref{Noetherdrift} is zero and the Noether current is preserved as the system evolves. Observe that conservation of this current is equivalent to the equation \eqref{redsohor1} which is the vertical equation of the second step of reduction that was obtained as an horizontal equation in the first reduction. On the other hand, if the first reduction is carried using the mechanical connection, the Noether current obtained by the invariance of the Lagrangian \eqref{redlagMECH} is
\begin{align*}
J_{2,\text{mech}}:T(\sS^1\times \sS^1 \times \sS^1)\oplus (\sS^1\times \sS^1 \times \sS^1\times\so(3)) \to &\R^3\\
(\theta,\dot{\theta},\xi)\mapsto &K(I+K)^{1}I\dot{\theta}.
\end{align*}
Now the action of the connection is not flat nor the action of $\sS^1\times \sS^1 \times \sS^1$ is horizontal, hence, the drift of the Noether current is for every $a\in\R^3$
\begin{align*}
\frac{d}{dt}\langle J(\theta,\dot{\theta},\xi),a\rangle&=-\langle \frac{\partial l_{1,O}}{\partial \xi}, \tilde{B}_{\theta}(\dot{\theta},a^{\sS^1\times \sS^1 \times \sS^1}_{\theta})+a_{\xi}^{\sS^1\times \sS^1 \times \sS^1\times\so (3)}\rangle\\
&=-\langle (I+K)\xi, ((I+K)^{-1}Ka)\times((I+K)^{-1}K\dot{\theta}-\xi)\rangle\\
&=\langle (I+K)^{-1}Ka,\Omega \times (I+K)\xi\rangle.
\end{align*}
This drift is equivalent to equation \eqref{redsover2} which again is the vertical equation of the second step of reduction that was obtained as an horizontal equation in the first one.

\section{Conclusion}\label{conclu}
\begin{figure}[h!]
\centering
\begin{tikzpicture}[scale=1.5]
\node (A) at (5,3.8) { $T(\sS^1\times \sS^1 \times \sS^1)\oplus \so(3) $};
\node (B) at (5,3.4) {Lagrangian $\ell_{1,O}$ and };
\node (C) at (5,3) {equations (\ref{redsover1}),(\ref{redsohor1})(MC connection)};
\node (D) at (5,2.6) {or equations (\ref{redsover2}),(\ref{redsohor2})(Mech. connection)};
\node (E) at (2.5,1.6) {$\so(3)\oplus  \R^3$};
\node (F) at (2.5,1.2) {Lagrangian $\ell$ and eqs. (\ref{EP1}),(\ref{EP2})};
\node (G) at (2.5,0.8) {Lagrangian $\ell_{2,S}$ and eqs. (\ref{redssover1}),(\ref{redssover2})};
\node (H) at (2.5,0.4) {Lagrangian $\ell_{2,O}$ and eqs. (\ref{redsosver}),(\ref{redsosvera})(MC connection)};
\node (I) at (2.5,0) { or eqs. (\ref{redsosver2}),(\ref{redsosvera2})(Mech. connection)};
\node (J) at (2.5,5) {$T(SO(3)\times \sS^1\times \sS^1 \times \sS^1)$};
\node (K) at (2.5,4.6) {Lagrangian $L$};
\node (L) at (0,3.4) {Lagrangian $\ell_{1,S}$ and };
\node (M) at (0,3) {equations (\ref{redsver}),(\ref{redshor})};
\node (N) at (0,3.8) { $TSO(3)\times \R^3 $};
\draw[->,font=\scriptsize,>=angle 90]
(K) edge node[above]{} (A)
(K) edge node[right]{} (N)
(K) edge node[right]{} (E)
(D) edge node[left]{} (E)
(M) edge node[left]{} (E);
\end{tikzpicture}
\caption{Diagram of the different reduction procedures performed in this paper and their equations}
\label{fig:sumup}
\end{figure}
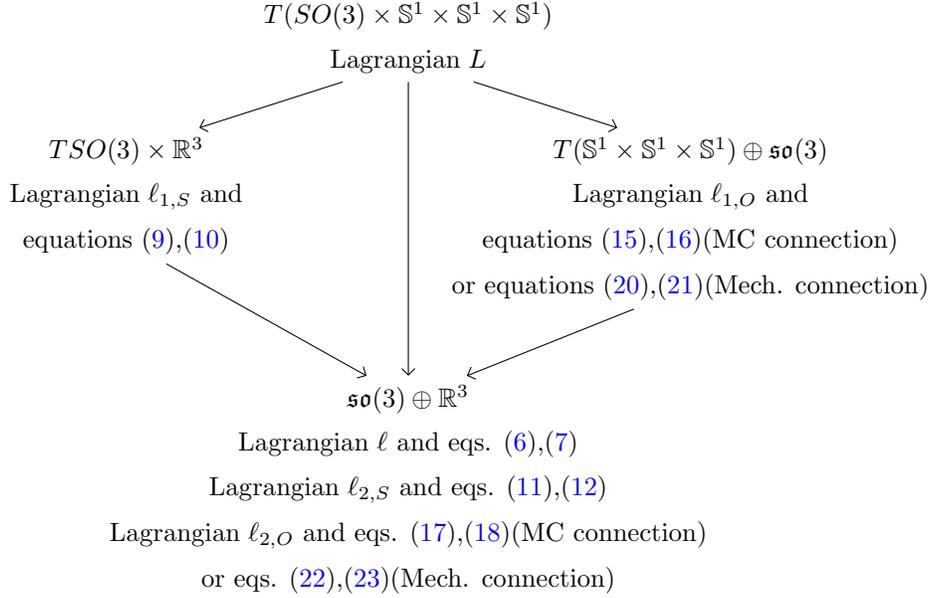

We have studied the free rigid body with rotors. In Section \ref{rigidbody} we obtain the equations of motion as Euler-Poincar\'e equations by reducing by the action of $SO(3)\times \sS^1\times \sS^1 \times \sS^1$ on itself. Section \ref{stagess} obtains equivalent equations by reducing first by $\sS^1\times \sS^1 \times \sS^1$ and then by $SO(3)$ showing that the free rigid body with rotors can be interpreted as a free rigid body with a force coming from the curvature term and also that the same equations can be though as horizontal an vertical equations or different factors of vertical equations. In section \ref{stagesso} it was carried out reduction by stages in the reverse order of section \ref{stagess}. This reduction is performed first using a trivial connection and then using the mechanical connection of the system. In the first case the Lagrangian is more complicated but the structure of $\mathfrak{LP}$-bundles involved is simple, while the mechanical connection simplifies the Lagrangian at the cost of complicating the $\mathfrak{LP}$-bundles involved. A diagram with the different reduction procedures performed on the rigid body in the aforementioned sections, as well as their equations, is shown in figure \ref{fig:sumup}. This mechanical system exemplifies the kind of calculations involved in Lagrangian reduction by stages and shows how this theory allows to think a problem from different viewpoints.

A field theoretical version of these techniques of reduction by stages can be found in \cite{FT}, where the example of a molecular strand consisting in a continuum of rigid bodies with rotors is explored. Further research may involve the study of these techniques for nonholonomic systems and their possible impact on the control of such systems like the spherical rolling robot.


\end{document}